%% file: paper.tex
\documentclass[submission,copyright,creativecommons]{eptcs}
\providecommand{\event}{ThEdu'18}
\usepackage{breakurl}             
\usepackage{underscore}           

\title{Towards Intuitive Reasoning in Axiomatic Geometry}
\author{Maximilian Dor\'e
\institute{Ludwig Maximilian University\\ Munich, Germany}
\email{m.dore@campus.lmu.de}
\and
Krysia Broda
\institute{Imperial College\\
London, UK}
\email{kb@imperial.ac.uk}
}
\def\titlerunning{Towards Intuitive Reasoning in Axiomatic Geometry}
\def\authorrunning{Maximilian Dor\'e \& Krysia Broda}

\usepackage{float} 
\usepackage{tikz}
\usetikzlibrary{positioning,shapes,arrows,matrix,fit,scopes,shapes.geometric}
\usepackage{color}
\usepackage{caption}
\usepackage{subcaption}
\usepackage{multicol}
\usepackage{lineno}
\usepackage{xparse}
\usepackage{amssymb} 
\usepackage{pifont}
\newcommand{\cmark}{\color{green}\ding{51}}%
\newcommand{\xmark}{\color{red}\ding{55}}%

\newcommand{\sref}[1]{Section {\ref{sec:#1}}}
\newcommand{\eref}[1]{Figure {\ref{text:#1}}}
\newcommand{\fref}[1]{Figure {\ref{fig:#1}}}
\newcommand{\ef}[1]{{\textsf{#1}}}
\newcommand{\cf}[1]{{\textcolor{blue}{#1}}}

\NewDocumentCommand{\etext}{sm}{
\begin{figure}[H]
\centering
\fbox{%
\begin{minipage}{.99\textwidth}
  \sffamily
    \IfBooleanT{#1}{
      \setcounter{linenumber}{1}
      \begin{internallinenumbers}
    }
      {#2}
    \IfBooleanT{#1}{
      \end{internallinenumbers}
    }
\end{minipage}}
\end{figure}
}
\newcommand{\ind}{\phantom{x}\hspace{3ex}}
\newcommand{\sind}{\phantom{x}\hspace{1.5ex}}

\tikzset{
  >= latex,
  el/.style={ellipse, draw, text width=8em, align=center},
  rs/.style={rectangle split, draw, rectangle split parts=#1},
  ou/.style={draw, inner xsep=1em, inner ysep=1ex, fit=#1},
  title/.style={font=\footnotesize, align=center, minimum width=15cm},
  proved/.style={line width=.3mm},
  typetag/.style={rectangle, draw=black!50, font=\footnotesize, anchor=west, minimum width=15cm, align=center},
  node distance=.5cm
}

\begin{document}
\maketitle
\begin{abstract}
  Proving lemmas in synthetic geometry is often a time-consuming endeavour
  since many intermediate lemmas need to be proven before interesting results
  can be obtained. Improvements in automated theorem provers (ATP) in recent
  years now mean they can prove many of these intermediate lemmas.
  The interactive theorem prover \textsc{Elfe} accepts mathematical texts
  written in fair English and verifies them with the help of ATP.
  Geometrical texts can thereby easily be formalized in \textsc{Elfe}, leaving
  only the cornerstones of a proof to be derived by the user. This allows for
  teaching axiomatic geometry to students without prior experience in formalized
  mathematics. 
\end{abstract}

\section{Introduction}

Formalizing mathematical proofs is something students usually do not apply or
even learn before their graduate studies.
Various tools for guiding proof construction in first order logic have been
proposed, for instance Pandora \cite{pandora} is an interactive  tool students can use to
construct correct natural deduction proofs, and more recently a Sequent calculus
trainer has been developed \cite{sequent}. A recent experimental project
is underway with first year students at Imperial College to use the
\textsc{Lean} prover
\cite{lean} as a vehicle for teaching both how to prove theorems and the use of
theorem provers to verify such proofs \cite{xena}. Another recent approach
teaches students mathematical reasoning by first introducing them to the proof assistant
 \textsc{Coq} \cite{coq}, and then transferring the skills they learned to informal proofs \cite{coqtextbook}.
While these successful projects create an understanding of mathematical
reasoning, none of them allows the user to write
proofs in English, as is done in typical undergraduate classes. Therefore, in
order to ease the introduction to
interactive theorem proving and constructing proofs, we have developed the \textsc{Elfe} system
\cite{csedu}. It has been previously used for proofs in discrete mathematics,
such as sets and relations, but recently, we have added a library to allow working within
synthetic geometry, which is the focus of this paper.

\begin{figure}[H]
\etext*{
Lemma: for all a,b,c,d,m. midpoint(m,b,c) and a-b-c and b-c-d and a-b $\equiv$ c-d and b $\neq$ c implies midpoint(m,a,d).\\
Proof: \\
\ind Assume midpoint(m,b,c) and a-b-c and b-c-d and a-b $\equiv$ c-d and b $\neq$ c.\\
\ind Then a-m $\equiv$ m-d since b-m $\equiv$ m-c and a-b $\equiv$ c-d.\\
\ind Note a-m-d: Then b-m-c by DefMidpoint. \\ \ind \ind Then a-b-m since a-b-c and
b-m-c. Then m-c-d since b-m-c and b-c-d.\\
\ind qed.\\
\ind Hence midpoint(m,a,d).\\
qed.
}
\caption{A simple proof in elementary geometry}
\label{text:example}
\end{figure}

Consider the exemplary \textsc{Elfe} text in \eref{example}.
The lemma in line 1 states that if a line has a midpoint \ef{m}, then we can extend the line on
both sides by the same distance such that the line between the outer points has
the same midpoint \ef{m}. In the statement, we use \ef{a-b-c} to denote that the
point \ef{b} lies in between the points \ef{a} and \ef{c}; the expression
\ef{a-b $\equiv$ c-d} expresses that the lines \ef{a-b} and \ef{c-d} are
equally long. The precise proposition of the lemma is therefore the following: given a line \ef{b-c} with
midpoint \ef{m}, adding points \ef{a} and \ef{d} on both sides of the line with an equal distance to \ef{b}
, respectively \ef{c}, will respect that \ef{m} is also the midpoint of the new
line between \ef{a-d}. The proof of the lemma is given in an intuitive way: We
observe in line 5 that the distance between \ef{a} and \ef{m} is the same as between
\ef{d} and \ef{m}. Furthermore, the point \ef{m} lies between \ef{a} and \ef{d}
as established in lines 6-8. We will understand in \sref{geometry} how the proof
works and how it is checked by the \textsc{Elfe} system.

Geometry is an attractive candidate for teaching formalized mathematics due to
its intuitive character --- even high school students can understand basic
lemmas involving parallel lines or midpoints. Consider the intuition behind the
previous proof, which is depicted below in \fref{simplegeoint}.
Since we extended the line between \ef{a} and \ef{c} on both sides by the same
length $\delta$, the new line \ef{a-d} must have the same midpoint as the line
\ef{b-c}. When transforming this kind of diagrammatic reasoning to an
axiomatic proof, the resulting mathematical text should be as close to the
informal proof as possible.

\begin{figure}[H]
\begin{center}
\begin{tikzpicture}[scale=0.8,every node/.style={draw=black,circle, fill=gray, inner sep=.04cm}]
    \node[label={a}] (a) at (0,0) {};
    \node[label={b}] (b) at (2,0) {} edge[dashed] (a);
    \node[label={m}] (m) at (5,0) {} edge (b);
    \node[label={c}] (c) at (8,0) {} edge (m);
    \node[label={d}] (d) at (10,0) {} edge[dashed] (c);

    \node[inner sep=0, color=white] (aD) at (-.04cm,.9) {};
    \node[inner sep=0, color=white] (bD) at (2.04cm,.9) {} edge[|-|, label={[xshift=-.7cm, yshift=-.01cm]$\delta$}] (aD);

    \node[inner sep=0, color=white] (cD) at (7.96cm,.9) {};
    \node[inner sep=0, color=white] (dD) at (10.04cm,.9) {} edge[|-|, label={[xshift=-.7cm, yshift=-.01cm]$\delta$}] (cD);
\end{tikzpicture}
\end{center}
\caption{The intuition behind the lemma from \eref{example}}
\label{fig:simplegeoint}
\end{figure}

The project \textsc{GeoCoq} \cite{geocoq} has
undertaken the effort to formalize a large body of elementary geometry
in the proof assistant \textsc{Coq}. The formalization makes use of different
axiomatizations of synthetic geometry, among these an axiom system thought out
by Alfred Tarski. The axiomatization is relatively straightforward and only involves
two basic predicates and around a dozen axioms. The proofs in \textsc{GeoCoq}
turn out to be quite long and complex. Many intermediate lemmas need to be
proven until interesting results can be obtained. The \textsc{Elfe} system in
contrast uses automated theorem provers (ATP) in the background 
to free its users from proving laborious steps. This approach has
turned out to be very useful in synthetic geometry, as steps in a proof that
are obvious to a human prover can be simply checked by the ATP in the background.
The user can therefore focus on the aspects of a proof that she wants to
understand better. The resulting proof texts are significantly shorter than the proofs in
\textsc{Coq}, which goes hand in hand with a reduced level of detail. This level
of detail is in some cases crucial, but when teaching students we believe that
a more high-level view of proofs can be beneficial.
We will see that the proof
style of \textsc{Elfe} is similar to the style of \textsc{Isar} \cite{isar}, a popular
language on top of the proof assistant \textsc{Isabelle} \cite{isabelle}; and shares some
resemblance with the \textsc{Mizar} system \cite{mizar}. In contrast to these systems, the
user can more freely choose proof paths and leave proof steps out since
\textsc{Elfe} utilizes the power of current ATP to check omitted details.

In the following, we will first introduce the concepts of formal
axiomatic geometry necessary for our purposes in \sref{background} before
further analyzing the above proof text in \sref{geometry}. We will see that the
system can also be used for more complex proofs in \sref{midpoint} before giving
an overview of related work in \sref{related} and ideas of further developments in \sref{outlook}.

\section{Background}
\label{sec:background}

We will first give a short overview of a first-order axiomatization of
elementary geometry in \sref{geo}, before turning to the theorem prover used in
this paper in \sref{elfe}.

\subsection{Axiomatic Geometry}
\label{sec:geo}

The endeavour of axiomatically capturing geometry was already pursued by Euclid
in his Elements. Hilbert and Tarski, among others, undertook the effort of giving
axiom systems for geometry in first-order logic. We will present the axiom
system of Tarski \cite{tarski} in the following.

\begin{figure}[H]
\etext*{\input{texts/tarski}}
\caption{Tarski's axioms in \textsc{Elfe}}
\label{text:axioms}
\end{figure}

\eref{axioms} presents the axiom system in the \textsc{Elfe} language. For now,
we do not need to know anything about the language except that it is a version of first-order logic.  
The only language feature we need to know about are \textit{notations}: Two new
notations for predicates are introduced in lines 1-2. The predicate
\ef{between(a,b,c)} expresses that three points \ef{a,b} and \ef{c}
are collinear, and that \ef{b} lies between the other two points. By introducing
the notation we can write \ef{a-b-c} instead of \ef{between(a,b,c)}.
Similarly, the notation \ef{a-b $\equiv$ c-d} stands for the predicate
\ef{equidistant(a,b,c,d)}. This predicate expresses that the line \ef{a-b} has the
same length as \ef{c-d}. These two predicates are sufficient to build a complete axiom system for
elementary geometry.
Note that at no point can we access the coordinates of a single point; instead
we will only talk about collinearities and lengths of lines in relation
to other lengths of lines. 

The first three axioms \ef{CongrRefl}, \ef{CongrIdent}, \ef{CongrTrans} specify
the behaviour of equidistance: a line \ef{a-b} is as long as its inverse definition \ef{b-a};
two points \ef{a} and \ef{b} collapse if the length of the line between them is
as long as the line \ef{c-c} for another point; and equidistance is transitive. 
The axiom \ef{SegmentConstr} is illustrated in \fref{segmentConstr}. It states that
we can extend a line \ef{a-b} by a point \ef{e}, with the line \ef{b-e} having
the same length $\delta$ as another line \ef{c-d}.

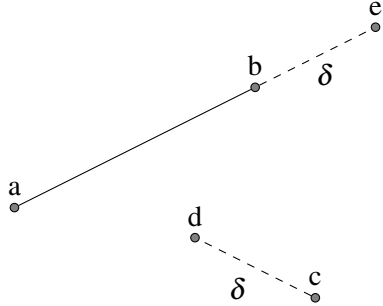
\begin{figure}[H]
\centering
\begin{tikzpicture}[scale=0.8,every node/.style={draw=black,circle, fill=gray, inner sep=.04cm}]
    \node[label={a}] (a) at (0,0) {};
    \node[label={b}] (b) at (4,2) {};
    \node[label={c}] (c) at (5,-1.5) {};    
    \node[label={d}] (d) at (3,-.5) {};
    \node[label={\color{white}e}, fill=white, draw=white] (f) at (6,3) {};
    \node[label={e}] (e) at (6,3) {};
    \path (a) edge (b);
    \path (c) edge[style=dashed] node[fill=white, draw=white, below left=.1cm and .05cm] {$\delta$} (d);
    \path (b) edge[style=dashed] node[fill=white, draw=white, below left=.05cm and -.31cm] {$\delta$} (e);
\end{tikzpicture}
\caption{The axiom \ef{SegmentConstr}}
\label{fig:segmentConstr}
\end{figure}

We will refer to \cite{schwabhauser} for an introduction to the other axioms. 
Note only that the axiom \ef{LowerDim} in line 11 asserts that there is
at least one proper triangle, and we therefore live at least in a 2-dimensional space.
Conversely, one can introduce an axiom to assert that there is no point outside
of the plane. For our purposes this is not necessary since the following
proof texts hold in arbitrary spaces with dimension greater or equal to 2.

The inspiration for our work on axiomatic geometry came from the project
\textsc{GeoCoq} \cite{geocoq}. \textsc{GeoCoq} attempts to formalize geometry in
\textsc{Coq}. \textsc{Coq} utilizes a dependently typed programming language to
formalize mathematics: via an extension of the
Curry-Howard correspondence, types of this language are understood as mathematical
propositions and programs of a type as proofs of the respective proposition.
The proofs in \textsc{GeoCoq} are therefore constructive, except of proofs
which require decidability of point equality, which is assumed as an axiom in the formalization.
\textsc{GeoCoq} explores several axiom systems, among them the axiom
system by Tarski introduced above. An excerpt of the resulting
\textsc{Coq} axiom set can be found in \eref{coqaxioms}. Since 'betweenness' is a
3-ary predicate, the type of \ef{Bet} maps three points to a truth-value,
i.e., \ef{Prop}. Similarly, \ef{Cong} maps four points to a truth-value. The
axioms such as \ef{cong_pseudo_reflexivity} are then straightforward versions of
the axioms we already got to know in \eref{axioms}. A current overview of the
status of \textsc{GeoCoq} can be found in \cite{geocoqarticle, geocoqarticle2}.

\begin{figure}[H]
\etext*{
Tpoint : Type;\\
Bet : Tpoint $\rightarrow$ Tpoint $\rightarrow$ Tpoint $\rightarrow$ Prop;\\
Cong : Tpoint $\rightarrow$ Tpoint $\rightarrow$ Tpoint $\rightarrow$ Tpoint $\rightarrow$ Prop;\\
cong_pseudo_reflexivity : forall A B, Cong A B B A;\\
cong_inner_transitivity : forall A B C D E F, Cong A B C D $\rightarrow$ Cong A B E F $\rightarrow$ Cong C D E F;\\
cong_identity : forall A B C, Cong A B C C $\rightarrow$ A = B;\\
segment_construction : forall A B C D,  exists E, Bet A B E $\wedge$ Cong B E C D;\\
...
}

\caption{Tarski's axioms in \textsc{GeoCoq} \cite{geocoq}}
\label{text:coqaxioms}
\end{figure}

\subsection{The \textsc{Elfe} Prover}
\label{sec:elfe}

After we already got to know how \textsc{Elfe} texts look we will sketch the
inner workings of the system. The \textsc{Elfe} system attempts at giving a
theorem prover that verifies mathematical proofs that are close to informal pen-and-paper
proofs. Its general mode of operation is inspired by the \textsc{System for
  Automated Deduction} (\textsc{SAD}) \cite{sad}. 

\begin{figure}[H]
\begin{center}
\tikzstyle{block} = [draw, fill=gray!15, minimum height=3em, minimum width=1.9cm, font=\footnotesize]
\tikzstyle{caption} = [draw,->,font=\scriptsize]
\begin{tikzpicture}[auto, node distance=2cm,>=latex']
    
    \node [block] (command) {Command line};
    \node [block, below=1cm of command] (web) {Web interface};

    \node [block, below right=.0cm and .5cm of command] (parser) {Parser};
    \node [block, right=1cm of parser] (verifier) {Verifier};
    
    \node [block, above right=.1cm and 1.98cm of verifier] (prover) {Prover(s)};
    \node [block, below=1cm of prover] (counter) {Countermodels};
    
    \draw [caption] (command) -- node {} (parser);    
    \draw [caption] (web) -- node {} (parser);
    \draw [caption] (parser) -- node {} (verifier); 
        
    \draw [caption] (verifier) edge [in=10,out=130] (command);
    \draw [caption] (verifier) edge [in=350,out=230] (web);
    
    \draw [caption] (verifier) -- node {} (prover);
    \draw [caption] (prover) -- node {} (verifier);
    \draw [caption] (verifier) -- node {} (counter);
    \draw [caption] (counter) -- node {} (verifier);
    
\end{tikzpicture}
\end{center}
\caption{Architecture of the \textsc{Elfe} system}
\label{fig:architecture}
\end{figure}
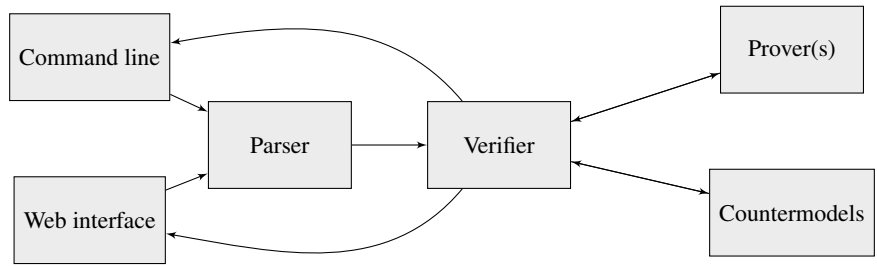

Proof texts can be entered via a command line interface or a web interface into
the system as depicted in \fref{architecture}. A text that should be checked is first parsed and transformed to a sequence of first-order
formulas. For example, the notation \ef{a-b-c} used in the initial proof will be
transformed to a first-order predicate $between(a,b,c)$. Notations like this are
not hard-coded in the system, but can be defined by the user in the proof text.
For example, after stating

\begin{center}
  \ef{Notation subset: A $\subseteq$ B.}
\end{center}
  
\noindent in the proof text, the user can simply write the Unicode sign $\subseteq$
between all kinds of sets. This allows for a rich input language that closely
resembles pen-and-paper mathematics.

After the syntactic sugar of notations is removed, the proof text is transformed
into a special data structure, the so-called \textit{statement sequence}. This intermediary data
structure implies certain proof obligations, which are checked by ATP in the
background. Some ATP such as \textsc{Vampire} \cite{vampire} or \textsc{E~Prover}~\cite{eprover} are called
in parallel; additionally, ATP are called to construct countermodels if an
obligation is wrong. The result of this verification process is then returned to
the user, either by telling them which derivations were correct or, if a
derivation step is incorrect, by presenting a countermodel. An in-depth
explanation of the \textsc{Elfe} system can be found in \cite{csedu}.

\section{Geometry in \textsc{Elfe}}
\label{sec:geometry}

With the axiom system given in \eref{axioms} it is possible to prove
interesting lemmas in elementary geometry. We will in the following build a
library on top of the axiom system and then analyze the introductory proof further.

The \textsc{Elfe} text depicted in \eref{geodefs} introduces definitions of important
geometrical concepts such as parallelism, solely built on top of the two
predicates \ef{between} and \ef{equidistant}. 
\begin{itemize}
\item The definition of \ef{DefCol} in line 1 generalizes the notion of betweenness:
  Three points are collinear if they lie on the same line, independent of the order of
  the points.
  \item A point \ef{m} is the midpoint of the line \ef{a-b} if it lies between \ef{a}
and \ef{b} and is equidistant to both points, as defined in \ef{DefMidpoint} in
lines 2-3. 
\item Since we are working in an arbitrary dimension greater than 1, four points do
not necessarily form a plane. The definition of \ef{DefCoplanar} in lines 4-5 characterizes
four points that do form a plane: If there is an intersection \ef{x} of two
lines formed by some combination of the four points \ef{a, b, c} or \ef{d}, then we
know that the four points lie in the same plane.
\end{itemize}

\begin{figure}[H]
\etext*{\input{texts/geodefs}}
\caption{Basic definitions upon Tarski's axioms}
\label{text:geodefs}
\end{figure}

Note that all these are explicit definitions, i.e., they do not extend the
theory built up by the axiom system in \eref{axioms}, but instead only combine
the two predicates used in the axioms. On top of these defined
predicates, we can build a notion of parallelism. \ef{DefParallelStrict} in
lines 7-8 captures the intuitive
understanding of parallelism: Two lines \ef{a-b} and \ef{c-d} are parallel
if they are coplanar and do not intersect, i.e., there exists no point \ef{x} that is collinear
with both lines.

When defining parallelism we also have to take into account a degenerate case:
Two lines can be parallel if they have an intersection, more precisely, if they
intersect completely and in fact describe the same line. This is captured in the
definition \ef{DefParallel} in lines 10-11: Two lines are parallel either if
they are strictly parallel, or all their points are collinear. In line 6, resp. 9,
we introduced additional notations such that we can write \ef{a-b$|-|$c-d} for
strict parallelism and \ef{a-b$||$c-d} for general parallelism.

With this geometry library in the background we can now further analyze the
exemplary proof of the introduction. Consider \eref{simplegeo}, which repeats the
introductory proof.

In line 1 the command \ef{Include} allows use of the geometry library.
All axioms and definitions previously stated are therefore in the context of our
proof text. We give the proposition stated by the lemma, and give it the name
\ef{MidpointExtension} in line 2. In order to prove the lemma, we assume the antecedent of
its main implication in line 5 and try to derive the consequent in line 12. The
lemma is universally quantified, fixing specific constants for the points in its
proof is left implicit, which is common practice in informal proofs. 

In order to show that \ef{m} is the midpoint of the line \ef{a-d}, we have to
show that the length of \ef{a-m} is equally long as \ef{m-d}, and that all three
points indeed lie on a line. The first statement is derived in line 6, it
directly follows from the axioms for equidistance and requires no further
proof. We emphasize that it follows from the two equidistances \ef{b-m $\equiv$
  m-c and a-b $\equiv$ c-d} with the keyword \ef{since}. Internally, the
statement after \ef{since} will be checked for validity by the ATP and, if correct,
be put in the context of the statement before the \ef{since}. 

The proof that all points lie on the same line requires more work. In line 7, we start a
subproof of the statement \ef{a-m-d} with the keyword \ef{Note}. All statements in lines 8-10 are then only
in the scope of this subproof; later in the proof in line 12 only the statement
\ef{a-m-d} remains in the context. With the construction \ef{by DefMidpoint} in
line 8 we limit the premises that are given to the background provers. Only the
definition of midpoint is required to derive the statement from the properties
of the constants. With this construction the user can speed up the search for
the background provers and ensure that he understands which premises make a
statement true. The derivations in line 9 and 10 do not limit the premises,
thus the whole context will be given to the ATP. If the background provers find
proofs for all proof obligations previously stated, the text is proven to
consist only of valid statements.

\begin{figure}[H]
\etext*{
  Include geometry.\\
Lemma MidpointExtension: for all a,b,c,d,m. midpoint(m,b,c) and a-b-c and b-c-d and a-b $\equiv$ c-d and b $\neq$ c implies midpoint(m,a,d).\\
Proof: \\
\ind Assume midpoint(m,b,c) and a-b-c and b-c-d and a-b $\equiv$ c-d and b $\neq$ c.\\
\ind Then a-m $\equiv$ m-d since b-m $\equiv$ m-c and a-b $\equiv$ c-d.\\
\ind Note a-m-d:\\
\ind \ind Then b-m-c by DefMidpoint.\\
\ind \ind Then a-b-m since a-b-c and b-m-c.\\
\ind \ind Then m-c-d since b-m-c and b-c-d.\\
\ind qed.\\
\ind Hence midpoint(m,a,d).\\
qed.
}
\caption{A simple proof in geometry}
\label{text:simplegeo}
\end{figure}

After removing the syntactic sugar of the notations, the text only consists of
structured first-order formulas, which is depicted in \eref{simplefol}.  

When the \textsc{Elfe} language is transformed to first-order logic, the system
internally keeps track of the variables in the proof. Since the user gave no
other information, the variables are assumed to be universally quantified, as
given in line 2. Inside the proof, the user fixes specific constants for the
variables. We have to take care of the change of variables to constants when
internally representing the structure of the proof, which we do below.

\begin{figure}[H]
\etext*{
...\\
Lemma: $\forall a,b,c,d,m. midpoint(m,b,c) \wedge between(a,b,c) \wedge
between(b,c,d) \wedge equidistant(a,b,c,d) \wedge b \neq c \rightarrow midpoint(m,a,d)$ \\
Proof: \\
\ind Assume $midpoint(m,b,c) \wedge between(a,b,c) \wedge between(b,c,d) \wedge
equidistant(a,b,c,d) \wedge b \neq c $. \\
\ind Then $equidistant(a,m,m,d)$ since $equidistant(b,m,m,c) \wedge equidistant(a,b,c,d)$.\\
\ind Note $between(a,m,d)$: \\
\ind \ind Then $between(b,m,c)$ by DefMidpoint. \\
\ind \ind Then $between(a,b,m)$ since $between(a,b,c) \wedge between(b,m,c)$.\\
\ind \ind Then $between(m,c,d)$ since $between(b,m,c) \wedge between(b,c,d)$. \\
\ind qed.\\
\ind Hence $midpoint(m,a,d)$. \\
qed.}
\caption{The proof in desugared first-order logic}
\label{text:simplefol}
\end{figure}

\begin{figure}[H]
\centering
\begin{center}
\begin{tikzpicture}  
\node (S) [title, label={[xshift=-7.5cm, yshift=0.05cm]\footnotesize $S$}] { $\forall a,b,c,d,m. midpoint(m,b,c) \wedge between(a,b,c) \wedge between(b,c,d) \wedge equidistant(a,b,c,d) \wedge b \neq c \rightarrow midpoint(m,a,d)$ };

    \node (S1) [below=of S, title, label={[xshift=-7.6cm, yshift=0.05cm]\footnotesize $S_1$}] {$midpoint(\cf{m},\cf{b},\cf{c}) \wedge between(\cf{a},\cf{b},\cf{c}) \wedge between(\cf{b},\cf{c},\cf{d}) \wedge equidistant(\cf{a},\cf{b},\cf{c},\cf{d}) \wedge \cf{b} \neq \cf{c} \rightarrow midpoint(\cf{m},\cf{a},\cf{d})$ };

        \node (S2) [below left=.8cm and .1cm of S1, typetag,  label={[xshift=-7.3cm, yshift=-.1cm, align=left]\footnotesize $S_2$}, draw=green!80, thick] {$midpoint(\cf{m},\cf{b},\cf{c}) \wedge between(\cf{a},\cf{b},\cf{c}) \wedge between(\cf{b},\cf{c},\cf{d}) \wedge equidistant(\cf{a},\cf{b},\cf{c},\cf{d}) \wedge \cf{b} \neq \cf{c}$ \\ \textsc{Assumed}};

        \node (S3) [below=.6cm of S2, title, label={[xshift=-7.4cm, yshift=0.05cm]\footnotesize $S_3$}] {$midpoint(\cf{m},\cf{a},\cf{d})$};

              \node (S4) [below=.4cm of S3, title, minimum width=14.7cm, label={[xshift=-7.3cm, yshift=0.05cm]\footnotesize $S_4$}] {$equidistant(\cf{a},\cf{m},\cf{m},\cf{d})$};

                  \node (S5) [below=1cm of S4.west, typetag, minimum width=14.7cm, label={[xshift=-7.1cm, yshift=-.1cm, align=left]\footnotesize $S_5$}, draw=orange!80, thick] {$equidistant(\cf{b},\cf{m},\cf{m},\cf{c}) \wedge equidistant(\cf{a},\cf{b},\cf{c},\cf{d})$ \\ \textsc{ByContext}};

                  \node (S6) [below=1.3cm of S5.west, typetag, minimum width=14.7cm, label={[xshift=-7.1cm, yshift=-.1cm, align=left]\footnotesize $S_6$}, draw=orange!80, thick] {$equidistant(\cf{a},\cf{m},\cf{m},\cf{d})$ \\ \textsc{ByContext}} edge[<-] (S5);

              \node (S4W) [draw=black!50, fit={(S4) (S5) (S6)}] {};

              \node (S7) [below=2.5cm of S4W.west, typetag, label={[xshift=-7.3cm, yshift=-.1cm]\footnotesize $S_7$},  draw=red, thick, dashed] {$between(\cf{a}, \cf{m}, \cf{d})$} edge[<-] (S4W);

              \node (S8) [below=1.3cm of S7.west, typetag,  label={[xshift=-7.3cm, yshift=-.1cm, align=left]\footnotesize $S_8$}, draw=orange!80, thick] {$midpoint(\cf{m}, \cf{a}, \cf{d})$ \\ \textsc{ByContext}} edge[<-] (S7);

         \node (S3W) [draw=black!50, fit={(S3) (S4W) (S7) (S8)}] {} edge[<-] (S2);

    \node (S1W) [draw=black!50, fit={(S1) (S2) (S3W)}] {};

\node [draw=black!50, fit={(S) (S1W)}] {};
\end{tikzpicture}
\end{center}
\caption{The statement sequence of \ef{MidpointExtension}}
\label{fig:statementSequence}
\end{figure}
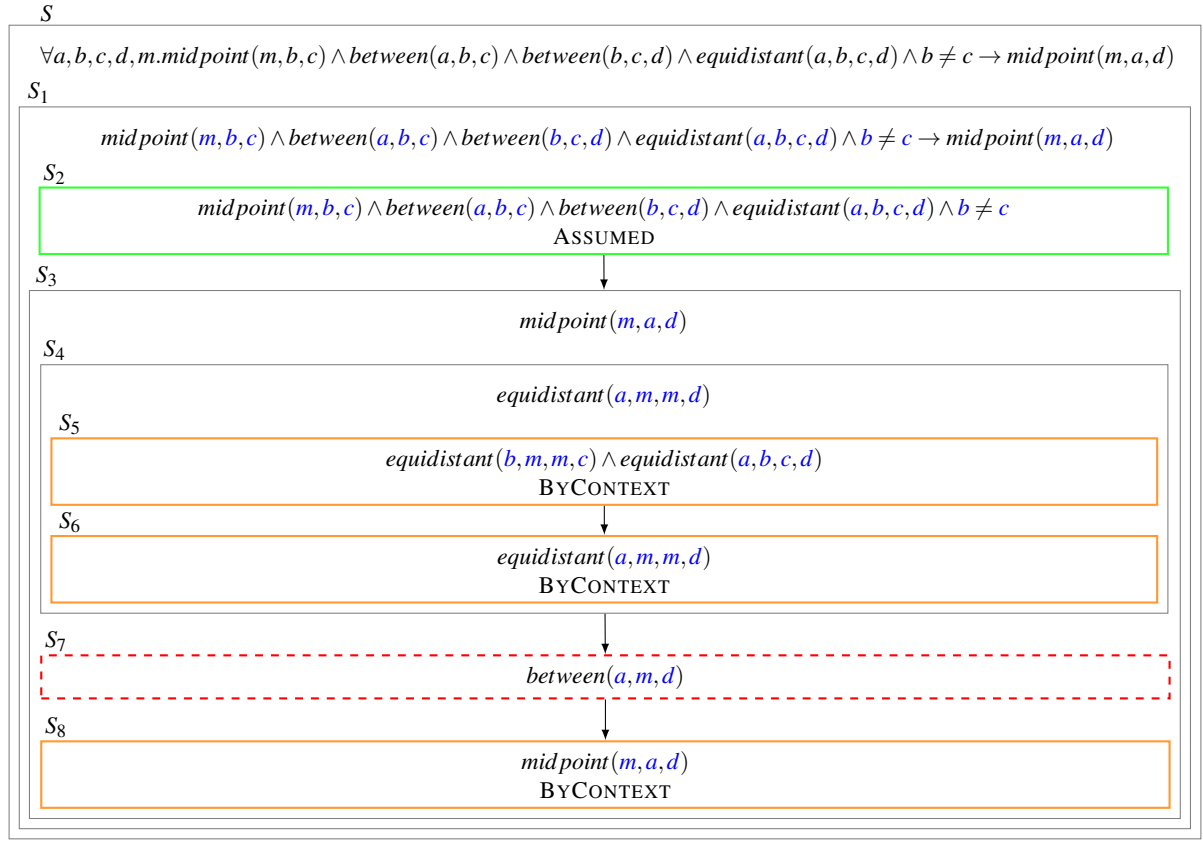

After the syntactic sugar is removed and the statements in an \textsc{Elfe} text
are in pure first-order logic, the text is transformed to an internal data
structure, the so-called \textit{statement sequences}, which provide
a basic representation of structured mathematical proofs. Consider
\fref{statementSequence}, which shows the statement sequence corresponding to
the proof of \ef{MidpointExtension}. A statement is represented with a box which
has a label in the upper left corner, e.g., the outer statement is called $S$.
Inside each statement is a mathematical sentence in first-order logic which is
called the \textit{goal} of the statement. For example, the goal of
statement $S$ is main statement of the lemma. Below the goal is the proof of the
goal, which can be of different kinds. In this instance, the proof of statement $S$ consists of
another statement, namely $S_1$.

The goal of $S_1$ is similar to the main lemma,
only that the universally quantified variables are fixed to constants, which are
coloured blue. One can easily see correctness of this construction by the
$\forall$-introduction rule of natural deduction: If a statement can be proven
with constants that are subject to no assumptions, the statement is also true for
universally quantified variables. The user did not have to explicitly state that
he proves the statement for a set of fixed constants, the system automatically
inferred this and created the statements accordingly.

The proof then employs a common proof tactic: In order to prove an
implication, the antecedent is assumed, additional derivations are made and finally, the
consequent follows. Accordingly, the proof of $S_1$ consists of the statement
sequence $S_2$ and $S_3$, which are connected with an arrow. This
illustrates that statement $S_2$ is in the context of $S_3$ and can be used in
its derivations. The proof of $S_2$ is simply \textsc{Assumed}, meaning that the
background provers can just take it as a premise in the following.

The proof of $S_3$ consists in turn of the statement sequence $S_4$, $S_7$ and
$S_8$. The statement $S_8$ repeats the goal of $S_3$ --- namely, that \ef{m} is
indeed a midpoint of line \ef{a-d}. In statement $S_8$ however, the proof is
\textsc{ByContext}: Its goal will be checked by the background
provers to see if they can derive it from the context. Crucially, the context
for statement $S_8$ consists of the goals of statements $S_2$, $S_4$ and $S_7$.
If we look back at the original proof text in \eref{simplegeo}, these statements
correspond to the following sentences: The assumption of line 5 and the
derivations of lines 6 and 7 are given to the background provers to see if
they can derive the sentence in line 12.

To sustain correctness, we of course also have to check if the sentences in
lines 6 and 7 are correct. In the statement sequence this is ensured by
requiring a proof sequence for both $S_4$ and $S_7$. The proof sequence of $S_4$
consists of two statements, which are both checked by the background provers.
Statement $S_5$ corresponds to the intermediary observation behind \ef{since} in the
original proof text. This statement is then in the context of $S_6$, which
checks the main statement of line 6.

The proof of statement $S_7$ consists similarly of several intermediary
derivations, we will omit it here. Note that these derivations, corresponding to
lines 8-10 in the proof text, are only local for proving $between(a,m,d)$.
Afterwards, only this statement is in the context for $S_8$. This allows for
scoping of derivation steps --- many steps are only relevant for a sub proof and the
background provers do not need to know about them subsequently.

\begin{figure}[H]
  \centering
    \includegraphics[width=1\textwidth]{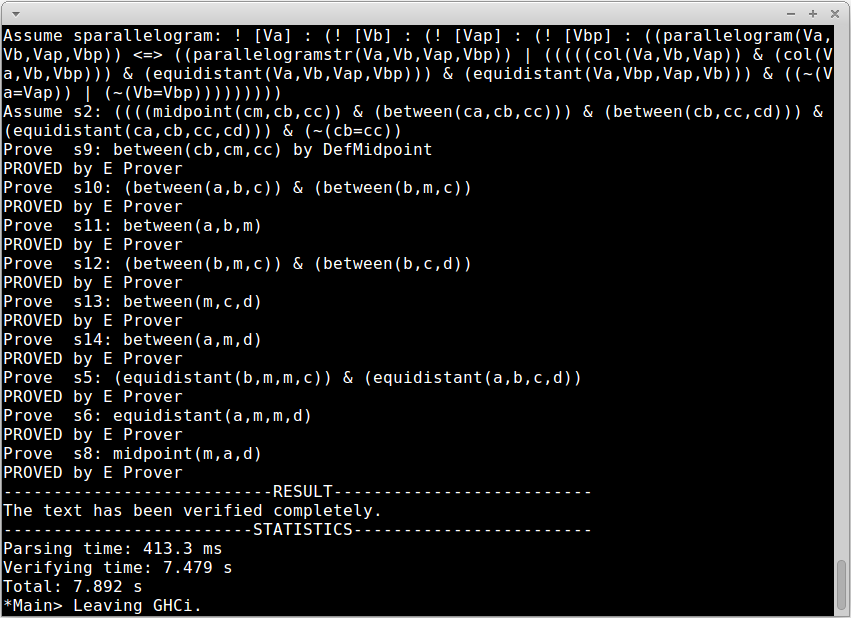}
    \caption{The terminal output when verifying \ef{MidpointExtension}}
    \label{fig:terminal}
\end{figure}

\fref{terminal} shows an excerpt of the terminal output when checking the
proof text from \eref{simplegeo}. In the first line, we can see one assumption
resulting from a definition in the background library, namely the definition of
parallelism. Below, statement $S_2$ is assumed, i.e., the antecedent of the
lemma. Underneath, all statements that were labelled with \textsc{ByContext} in
\fref{statementSequence} are checked by the background provers. For example, in
the last line before the result we can see that statement $S_8$ was proved by
\textsc{E Prover}. Below the verification process, the overall result of the
verification process is summarized. Since the text is correct and all proof
obligations could be verified by the ATP, the result is positive. If the user
made a mistake or no proof was found for an obligation, the statement and its
context is printed. Finally, some statistics of the verification process are given.

The web interface can be used for a more intuitive access to the \textsc{Elfe}
system. \fref{web} shows the web interface of the prover after checking the
lemma \ef{MidpointExtension}. Since all lines are coloured green, the system
has accepted the proof. The user can then inspect the verification process by
setting the focus on a line. In the screenshot, the user has selected line 7.
Below the input field we can see more information about the sentence: Two
obligations have been checked by the background provers, in both cases \textsc{E
Prover} has found a proof. The first obligation corresponds to the main
statement of the sentence, the second obligation to the intermediary derivation
step after the \ef{since}. The user can see the internal representation of both
statements in first-order logic, after removing the syntactic sugar from notations.
An instance of the web interface can be found online.\footnote{\url{https://elfe-prover.org/} \\ In order to inspect
  the introductory proof go to:
  \url{https://elfe-prover.org/?example=Midpoint\%20Extension} \\ Click on "Verify"
                                on the right to check the lemma and set the
                                focus in the text to inspect the verification
                                process.}

\begin{figure}[H]
  \centering
    \includegraphics[width=1\textwidth]{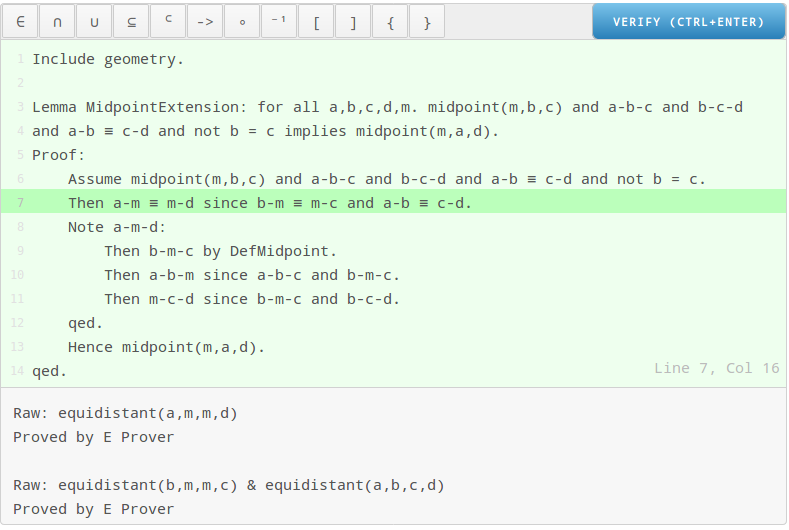}
    \caption{The web interface of \textsc{Elfe}}
    \label{fig:web}
\end{figure}

\section{A Midpoint Theorem}
\label{sec:midpoint}

We will now turn to proving a more intricate lemma. Consider the following statement: 

\ef{for all a,b,a',b',m. a $\neq$ b and midpoint(m,a,a') and midpoint(m,b,b') implies a-b$||$a'-b'.}

The \textsc{Elfe} sentence states that if two lines \ef{a-a'} and \ef{b-b'}
intersect in a point \ef{m}, which is also a midpoint of both lines, then the lines
\ef{a-b} and \ef{a'-b'} must be parallel to each other.

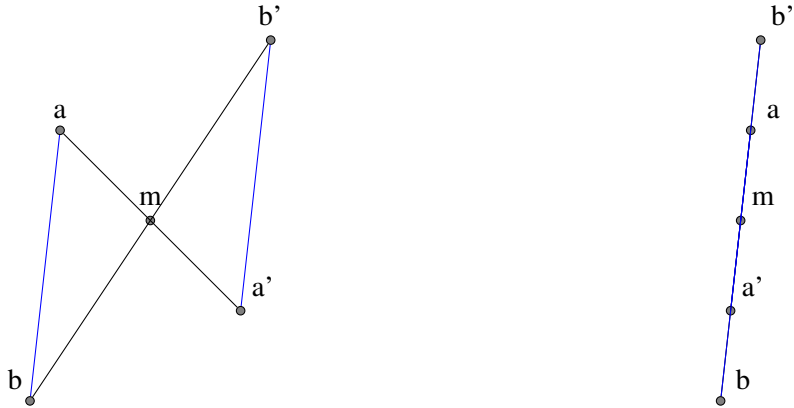
\begin{figure}[H]
\center
\begin{multicols}{2}
\begin{tikzpicture}[scale=0.8,every node/.style={draw=black,circle, fill=gray, inner sep=.04cm}]
    \node[label={a}] (a) at (.5,4.5) {};
    \node[label={[xshift=-.2cm, yshift=0cm]b}] (b) at (0,0) {};
    \node[label={m}] (m) at (2,3) {};
    \node[label={[xshift=.3cm, yshift=0cm]a'}] (ap) at (3.5,1.5) {};    
    \node[label={b'}] (bp) at (4,6) {};
    \path (a) edge (ap);
    \path (b) edge (bp);
    \path (a) edge[color=blue] (b);
    \path (ap) edge[color=blue] (bp);
\end{tikzpicture}\\
  \begin{tikzpicture}[scale=0.8,every node/.style={draw=black,circle, fill=gray, inner sep=.04cm}]
    \node[label={[xshift=.3cm, yshift=0cm]a}] (a) at (.5,4.5) {};
    \node[label={[xshift=.3cm, yshift=0cm]b}] (b) at (0,0) {};
    \node[label={[xshift=.3cm, yshift=0cm]m}] (m) at (.333,3) {};
    \node[label={[xshift=.3cm, yshift=0cm]a'}] (ap) at (0.166,1.5) {};
    \node[label={[xshift=.3cm, yshift=0cm]b'}] (bp) at (.666,6) {};
    \path (a) edge (ap);
    \path (b) edge (bp);
    \path (a) edge[color=blue] (b);
    \path (ap) edge[color=blue] (bp);
  \end{tikzpicture}
\end{multicols}
\caption{Parallelism of lines \ef{a-b} and \ef{a'-b'}}
\label{fig:cases}
\end{figure}

Consider the left illustration in \fref{cases}: Lines \ef{a-a'} and \ef{b-b'} cross each other in
their respective midpoint, the lines \ef{a-b} and \ef{a'-b'}, coloured blue,
are therefore parallel. When proving the lemma, we also have to consider a degenerate case, which is
shown on the right hand side of \fref{cases}: If both lines \ef{a-a'} and \ef{a'-b'}
lie on top of each other, then all lines formed by the points are parallel to
each other, in particular \ef{a-b} and \ef{a'-b'}.

\begin{figure}[H]
\centering
\fbox{%
\begin{minipage}{.99\textwidth}
  \sffamily
      \setcounter{linenumber}{1}
      \begin{internallinenumbers}
Lemma: for all a,b,a',b',m. a $\neq$ b and midpoint(m,a,a') and midpoint(m,b,b') implies a-b$||$a'-b'.\\
Proof: \\
\ind Assume a $\neq$ b and midpoint(m,a,a') and midpoint(m,b,b').\\
\ind Case col(a,b,b'):\\
\ind \ind Then a' $\neq$ b' and col(a,a',b') and col(b,a',b') by MidpointCol. \\
\ind \ind Then a-b$||$a'-b' by DefParallel.\\
\ind qed.\\
\ind Case not col(a,b,b'):
\end{internallinenumbers}
\vspace{-1.1em}
\ind \ind ... \\
\setcounter{linenumber}{33}
\begin{internallinenumbers}
\ind qed.\\
\ind Hence a-b$||$a'-b'.\\
qed
      \end{internallinenumbers}
\end{minipage}}
\caption{The two main cases of the lemma}
\label{text:cases}
\end{figure}

We will now prove the above statement. Consider \eref{cases}. In line 1 we state
the lemma and give its proof below. We assume the left hand side of the
statement in line 3 and want to derive the conclusion in line 35. Since we have
to take into account the degenerate case, we will introduce a case distinction:

If points \ef{a, b} and \ef{b'} are collinear, we are in the degenerate case. The
proof of this case is given in lines 5-6: We use the following lemma to derive
that also \ef{a,a'} and \ef{b'} and respectively \ef{a',b'} and \ef{b} must be collinear:

\ef{Lemma MidpointCol: for all a,b,a',b',m. a $\neq$ b and midpoint(m,a,a') and
  midpoint(m,b,b') and col(a,b,b') implies a' $\neq$ b' and col(a,a',b') and
  col(b,a',b').}

We will not give a proof of the lemma \ef{MidpointCol} here. This proves the
degenerate case of the definition \ef{DefParallel}.  

\begin{figure}[H]
\centering
\fbox{%
\begin{minipage}{.99\textwidth}
  \sffamily
      \setcounter{linenumber}{1}
      \begin{internallinenumbers}
\setcounter{linenumber}{8}
Case not col(a,b,b'):   \\
\ind Note a' $\neq$ b':\\
\ind \ind Assume a' = b'. \\
\ind \ind Then a'-b'-m and m-a' $\equiv$ m-b'.\\
\ind \ind Then m-a-b and m-a $\equiv$ m-b. \\ 
\ind \ind Then a = b by BetweenCong. \\
\ind \ind Hence contradiction. \\
\ind qed.\\
\ind Note coplanar(a,b,a',b'):\\
\ind \ind Then a-m-a' and b-m-b' by DefMidpoint. \\
\ind \ind Then col(a,a',m) and col(b,b',m) by ColPerm, DefCol.\\
\ind \ind Then coplanar(a,b,a',b') by DefCoplanar. \\
\ind qed. \\
\ind Note not exists x. col(x,a,b) and col(x,a',b'):
\end{internallinenumbers}
\vspace{-1.1em}
\ind \ind ... \\
\setcounter{linenumber}{30}
\begin{internallinenumbers}
\ind qed. \\
\ind Then a-b$|-|$a'-b' by DefParallelStrict.\\
\ind Then a-b$||$a'-b' by DefParallel.\\
qed.
      \end{internallinenumbers}
\end{minipage}}
\caption{Strict parallelism in the non-degenerate case}
\label{text:notes}
\end{figure}

In order to prove the non-degenerate case, i.e., if \ef{a,b} and \ef{b'} are not
collinear, we have to give a more intricate proof, its sketch is given in
\eref{notes}. We will have to prove that both lines are strictly parallel, the
definition of strict parallelism as given in \eref{geodefs} requires that we
prove three properties:

\begin{itemize}

\item \ef{a $\neq$ b} and \ef{a' $\neq$ b'}: The first inequality already holds by
assumption, we thus only need to prove the latter inequality. This is done in
lines 10-14. We assume that \ef{a'} is equal to \ef{b'} and want to derive a
contradiction. We will use another auxiliary lemma for this:

\ef{Lemma BetweenCong: for all a,b,c. a-b-c and a-b $\equiv$ a-c implies b = c.}

This lemma states that if we have a line \ef{a-b-c} and the distance between
\ef{a} and \ef{b} respectively \ef{c} is the same, then \ef{a} must be equal to
\ef{b}. A proof of this lemma is given in \cite{springer}. We can make use of it in line 13 by unifying \ef{c} with \ef{m},
which is given in line 12. This statement in turn follows from the fact that we
also have \ef{a'-b'-m and m-a' $\equiv$ m-b'}, and \ef{a} is \ef{a'} mirrored on
\ef{m}, and \ef{b} is \ef{b'} mirrored on \ef{m}.

\item All points are coplanar: To prove that all points lie in the same plane we have
to find an intersection of two lines formed by a combination of all four points.
This witness of coplanarity is conveniently given by the midpoint \ef{m}. By the
definition of a midpoint \ef{m} must lie between \ef{a} and \ef{a'},
respectively \ef{b} and \ef{b'} as stated in line 17. Then in particular the
weaker notion of collinearity holds and we can can conclude that all four points
are collinear in line 19.

\item The remaining property we need to prove is that both lines \ef{a-b} and
\ef{a'-b'} have no intersection, which we will do in the following.

\end{itemize}

\begin{figure}[H]
\centering
\fbox{%
\begin{minipage}{.99\textwidth}
  \sffamily
      \setcounter{linenumber}{1}
      \begin{internallinenumbers}
\setcounter{linenumber}{21}
Note not exists x. col(x,a,b) and col(x,a',b'):\\
\ind Assume exists x. col(x,a,b) and col(x,a',b').\\
\ind Take x such that col(x,a,b) and col(x,a',b').\\
\ind Take x' such that x-m-x' and m-x' $\equiv$ m-x by SegmentConstr.\\
\ind Then col(a,b,x') and col(a',b',x'). \\
\ind Then col(b',x,x') since col(b',a',x) and col(b',a',x').\\
\ind  Then col(b,x,x') since col(b,a,x) and col(b,a,x'). \\
\ind Then col(b,x,b') since col(b,x,x') and col(b',x,x'). \\
\ind Then col(a,b,b') since col(b,x,b') and col(b,x,a). \\
\ind Hence contradiction.\\ 
qed.
      \end{internallinenumbers}
\end{minipage}}
\caption{Proving there is no intersection of both lines}
\label{text:nondeg}
\end{figure}

You can find the corresponding \textsc{Elfe} text in
\eref{nondeg}. We suppose that there is one intersection, and will derive a
contradiction from that.

\begin{figure}[H]
\begin{center}
\begin{tikzpicture}[scale=0.6,every node/.style={draw=black,circle, fill=gray, inner sep=.04cm}]
    \node[label={a}] (a) at (.5,4.5) {};
    \node[label={b}] (b) at (0,0) {};
    \node[label={b'}] (bp) at (4,6) {};
    \node[label={m}] (m) at (2,3) {};
    \node[label={a'}] (ap) at (3.5,1.5) {};    
    \path (a) edge (ap);
    \path (b) edge (bp);
    \path (a) edge[color=blue] (b);
    \path (ap) edge[color=blue] (bp);
    \node[label={x}] (x) at (1.5, -1.5) {};
    \path (b) edge[style=dashed, bend right] (x);
    \path (ap) edge[style=dashed, bend left] (x);
    \node[label={x'}] (xp) at (2.5, 7.5) {};
    \path (x) edge (xp);
    \path (bp) edge[style=dashed, bend right] (xp);
    \path (a) edge[style=dashed, bend left] (xp);
  \end{tikzpicture}
\end{center}
  \caption{Deriving a contradiction}
  \label{fig:contrad}
\end{figure}
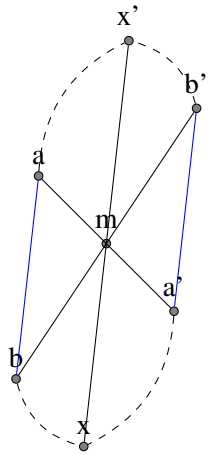

In order to derive a contradiction, we first fix a point \ef{x} that
evinces both collinearities with \ef{a-b} and \ef{a'-b'} in line 23. The supposed
situation is depicted in \fref{contrad}. We have to draw curves to represent all
collinearities --- this already suggests that we are indeed deriving a contradiction.
In line 24 we then employ the axiom \ef{SegmentConstr} to construct a point \ef{x'}
that is \ef{x} mirrored on \ef{m}.

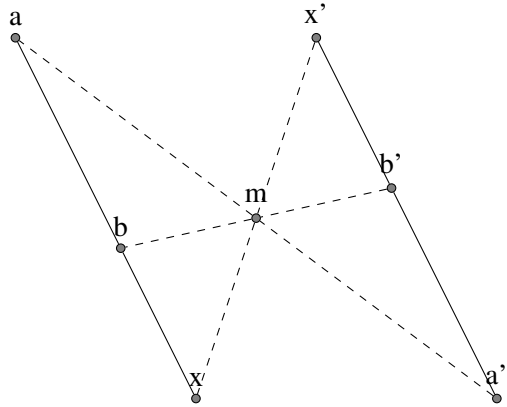
\begin{figure}[H]
\begin{center}
\begin{tikzpicture}[scale=0.8,every node/.style={draw=black,circle, fill=gray, inner sep=.04cm}]
    \node[label={a}] (a) at (0,6) {};
    \node[label={b}] (b) at (1.75,2.5) {} edge (a);
    \node[label={x}] (x) at (3,0) {} edge (b);
    \node[label={m}] (m) at (4,3) {} edge[dashed] (a) edge[dashed] (b) edge[dashed] (x);
    \node[label={a'}] (ap) at (8,0) {} edge[dashed] (m);
    \node[label={b'}] (bp) at (6.25,3.5) {} edge[dashed] (m) edge (ap);
    \node[label={x'}] (xp) at (5,6) {} edge[dashed] (m) edge (bp);
\end{tikzpicture}
\end{center}
\caption{Mirroring a line around a midpoint}
\label{fig:mirror}
\end{figure}

In line 25 we observe that the point \ef{x'} must lie on the same line as \ef{a-b} and
respectively \ef{a'-b'}. This is the case since we have \ef{col(a',b',x)}
respectively \ef{col(a,b,x)} and then can mirror all three points on a common
point \ef{m} to derive \ef{col(a,b,x') and col(a',b',x')}. The illustration in \fref{mirror} demonstrates the
intuitive reasoning behind this observation. In our proof text we do not have to give more derivation steps
since the background provers are able to derive all collinearities.

In lines 26-29 we employ the following lemma four times:

\ef{Lemma ColTrans: for all a,b,c,d. (not a = b) and col(a,b,c) and col(a,b,d) implies col(a,c,d).}

In line 29 we therefore have \ef{a}, \ef{b} and \ef{b'} are collinear, which
contradicts our initial assumption of the non-degenerate case, namely that these
points form a proper triangle. This completes our proof of the lemma.

\section{Related Work}
\label{sec:related}

The venture of using formal systems for teaching mathematics has been employed
manifold, e.g., the \textsc{Pandora} system \cite{pandora} included an automated
tutor that students could ask for help if they were stuck. Another project
\cite{coqtextbook} uses the \textsc{Coq} proof assistant in an interesting way.
When teaching students, they keep strictly to the \textsc{Coq} syntax in the first step.
Only in the second step they encourage students to write English comments
detailing the steps being performed, but in the manner of an ordinary textbook
proof. Interestingly, this is a different perspective from the one we take in
\textsc{Elfe} --- we jump in straight away with a textbook proof, gradually
building up notation as students become more familiar.

This is due to the fact that the \textsc{Elfe} system is the product of a
different line of research. We have already mentioned \textsc{SAD} \cite{sad}. \textsc{SAD}
works similarly to \textsc{Elfe} since it takes as input mathematical texts in a language fairly
close to natural language before checking them with the help of ATP. The primary
aim of \textsc{SAD} was to provide a text verifier in particular for
mathematical researchers. \textsc{Elfe} uses the same mode of operation, but
focuses with its interface and axiomatizations on mathematical beginners.

\textsc{SAD} in turn was influenced by other systems, in particular, the
\textsc{Mizar} \cite{mizar} system and the \textsc{Isabelle} \cite{isabelle} prover.
\textsc{Mizar} was already developed as soon as 1973 and is still under active
development. The first version of \textsc{Isabelle} dates back to 1986. The
\textsc{Isar} language, which aims at giving an intuitive proof language for
\textsc{Isabelle}, was developed in 1999 \cite{isar}. Since then, \textsc{Isar}
has become the standard for mathematical texts in \textsc{Isabelle}.

In the following, we will put \textsc{Elfe} in the context of the development of
other proof assistants in \sref{related1}. We then take a closer look at the
proof style of \textsc{Elfe} in comparison with \textsc{Mizar} and
\textsc{Isabelle} in \sref{related2}.

\subsection{Comparison of Proof Assistants}
\label{sec:related1}

Table \ref{tab:comp} depicts a comparison of several popular theorem provers and
the \textsc{Elfe} system. We can differentiate between two main lines of
provers: \textsc{Mizar} influenced the development of \textsc{Isabelle},
\textsc{SAD} and \textsc{Elfe}. Another tradition builds on the Curry-Howard
correspondence that interprets types as propositions and programs of a type as
their proof.  The prover \textsc{Coq} \cite{coq} builds on top of the Calculus of Constructions
, a dependently typed
programming language. \textsc{Lean} \cite{lean} and \textsc{Agda} \cite{agda} present further
elaborations of that logical system and have been developed in the last decade.

\begin{figure}[H]
\centering
\begin{tabular}{l|ccccccc}
                    & \textsc{Mizar} & \textsc{Isabelle} & \textsc{Coq} & \textsc{Lean} & \textsc{Agda} & \textsc{SAD} & \textsc{Elfe}  \\ \hline \hline
Logic               &   SOL   &   HOL  &  HOL   &   HOL   &   HOL   &   FOL  & FOL \\
Typing               &   ++    &   ++   &  +++  &  +++  &  +++  &  +  & + \\
ATP for proof search          &    \cmark   &  \cmark       &  \xmark   &  \xmark  &   \xmark   &  \cmark   & \cmark \\
Premise annotation         &   necc.   &  necc.        &  necc.   &   necc.    &   necc.   & not possible & possible  \\
Reasoning capabilities            &   \cmark    &    \cmark    & \cmark    &  \cmark    &  \cmark    &   \cmark  & \xmark \\
Unicode          &   \xmark   &     \cmark    &   \cmark  &   \cmark   &   \cmark   &   \xmark  & \cmark 
\end{tabular}
\caption{Comparison of different provers}
\label{tab:comp}
\end{figure}

As shown in the first line, \textsc{Isabelle} uses a higher-order logic with
simple types. The dependently typed provers \textsc{Coq}, \textsc{Lean} and
\textsc{Agda} have a more expressive type system that employs dependent types,
i.e., types that depend on values. \textsc{SAD} and \textsc{Elfe} use
first-order logic internally. \textsc{Mizar} adds typing on top of
second-order logic, which can mostly be represented with predicates of
first-order logic. This type system allows for a simple notion of dependent types, but
its expressive power is not comparable to dependently typed systems in the spirit
of \textsc{Coq}. \textsc{Elfe} has also a
primitive version of typing, as variables can be defined to fall under a certain
predicate and afterwards be used without further annotation.

Automated theorem provers can be employed for proof search in \textsc{Mizar}, and in
\textsc{Isabelle} via its extension \textsc{Sledgehammer}. \textsc{SAD} and
\textsc{Elfe} trust the results of the ATP and do not
need annotations in the proof. On the other hand, \textsc{Isabelle} and \textsc{Mizar}
translate proofs generated by ATP in their own language that is then checked
with their own reasoning mechanisms. Most systems, including \textsc{Elfe}, allow for
using Unicode characters for mathematical notations.

\subsection{\textsc{Mizar}, \textsc{Isabelle} and \textsc{Elfe} Compared}
\label{sec:related2}

As we have seen in the previous section, \textsc{Elfe} stands in the history of
\textsc{Mizar} and \textsc{Isabelle}. We will take a look at an exemplary proof
in the different systems.

Consider \fref{simpleElfe}, which depicts two proofs of the same lemma in
\textsc{Elfe}. The proposition of the lemma is a simple observation about the
betweenness relation.

\begin{figure}[H]
  \begin{subfigure}[t]{0.49\textwidth}
\etext{
    \small
Lemma: for all a,b,c,d. a-b-d and b-c-d implies a-b-c. \\
Proof: \\
\ind Assume a-b-d and b-c-d.\\
\ind Take x such that b-x-b and c-x-a by Pasch.\\
\ind Then b = x since b-x-b.\\
\ind Then c-b-a since c-x-a and b = x.\\
\ind Hence a-b-c.\\
qed.
}
\end{subfigure}
\begin{subfigure}[t]{0.02\textwidth}
\end{subfigure}
\begin{subfigure}[t]{0.49\textwidth}
\etext{
    \small
Lemma: for all a,b,c,d. a-b-d and b-c-d implies a-b-c. \\
Proof: \\
\ind Assume a-b-d and b-c-d.\\
\ind Take x such that b-x-b and c-x-a.\\
\ind Hence a-b-c.\\
qed.
}
\end{subfigure}
\caption{Two possible proofs of the same lemma in \textsc{Elfe}}
\label{fig:simpleElfe}
\end{figure}

The first proof in \textsc{Elfe} is very similar to the proofs in \textsc{Mizar}
and \textsc{Isabelle} depicted in \fref{mizarIsabelle}, which have been
constructed by \cite{grabowski} and \cite{isabelleTarski}.
\textsc{Elfe} also allows for other proofs, such as the second proof in
\fref{simpleElfe}. Here, we only need to observe that we can constrcut a point
\ef{x} with appropriate properties to show the lemma, we also do not need to
state that the \ef{Pasch} axiom is needed for that derivation. While this text
is not necessarily instructive, it can be used by a student working on the text
as first step in constructing a more verbose proof. The student can
step by step refine the proof, whereas in \textsc{Mizar} and \textsc{Isabelle}
it is not immediately obvious if the chosen proof path will succeed. 

\begin{figure}[H]
  \begin{subfigure}[t]{0.49\textwidth}
\etext{
    \small
theorem LineExtension: \\
\sind for S being TarskiGeometryStruct\\
\sind for a, b, c, d being POINT of S\\
\sind st between a,b,d \& between b,c,d holds\\
\sind between a,b,c\\
proof\\
\sind let S be TarskiGeometryStruct ; :: thesis:\\
\sind let a, b, c, d be POINT of S; :: thesis:\\
\sind assume H1: between a,b,d ; :: thesis:\\
\sind assume between b,c,d ; :: thesis:\\
\sind then consider x being POINT of S such that\\
\sind X1: ( between b,x,b \& between c,x,a ) by H1, A7;\\
\sind b = x by X1, A6;\\
\sind hence between a,b,c by Bsymmetry, X1; :: thesis:\\
end;
}
\end{subfigure}
\begin{subfigure}[t]{0.02\textwidth}
\end{subfigure}
\begin{subfigure}[t]{0.49\textwidth}
\etext{
    \small
theorem line_extension:\\
\sind assumes "B a b d" and "B b c d"\\
\sind shows "B a b c"\\
proof -\\
\sind from `B a b d` and `B b c d` and A7' [of a b d b c]\\
\sind obtain x where "B b x b" and "B c x a" by auto\\
\sind from `B b x b` have "b = x" by (rule A6')\\
\sind with `B c x a` have "B c b a" by simp\\
\sind thus "B a b c" by (rule th3_2)\\
qed
}
\end{subfigure}
\caption{The proof in \textsc{Mizar} \cite{grabowski} and \textsc{Isar/Isabelle}
\cite{isabelleTarski}}
\label{fig:mizarIsabelle}
\end{figure}

The proof texts in
\textsc{Elfe}, \textsc{Mizar} and \textsc{Isabelle} for the simple lemma are all
quite legible and provide insight into why the proof succeeds. The proof texts of
the latter two provers seem more technical, which is partly owed to the fact
that the systems are more sophisticated and allow for annotating proof steps.
Proofs in \textsc{Elfe} rely on ATP and can therefore be shorter than the other
proofs. It is at the discretion of the user to decide when he considers a proof
complete. Thereby, the \textsc{Elfe} system allows for more freely exploring a
theory generated by an axiom system.

The \textsc{Elfe} system is implemented in a lean code base and allows for
simple extensions --- in particular, new proof structures and language
constructs can simply be added to the sytem if they can be represented with statement sequences. Background
provers can also simply be integrated by changing the configuration of the
system with a single line.

New axiom systems can be easily created since no background mathematical system
is assumed. When using existing libraries, it is not necessary to memorize
possible premises since all can be used by the ATP to find proof for a
derivation step.

By focusing on mathematical beginners as users, the prover can
be used as a valuable didactic device, and thus we think that the \textsc{Elfe} system fills a gap in the zoo of proof assistants.
The system certainly has not reached the
level of sophistication of \textsc{Isabelle}, but this allows the system to be
used as a test bed for lean axiomatizations and highly legible proof
texts.

\section{Outlook}
\label{sec:outlook}

In this paper we have seen how one can formally prove complex lemmas in
elementary geometry with the help of the interactive theorem prover
\textsc{Elfe}. 
By giving most of the low-level proof work to automated theorem
provers, the resulting proof texts  are concise and give a good overview of the
main proof ideas, which makes \textsc{Elfe} proofs more similar to informal
pen-and-paper proofs.  
In order to more deeply investigate different axiom systems, the lack of detail in
\textsc{Elfe} proofs might be hindering since it is not necessary to specify
which premises are used, and where, in a proof. It is therefore not always apparent
which exact constructions make a derivation true.

On top of the already present \textsc{Elfe} libraries for working with sets,
relations and functions, it might be interesting to investigate other domains
in discrete mathematics such as graph theory. Synthetic characterizations of
non-discrete domains like topology make them also possible candidates for a formalization in
\textsc{Elfe}.

Geometry is a great show case for formalized mathematics due to its illustrative
power --- its statements can be understood already by high-school students.
Geometrical proof texts in the \textsc{Elfe} system can provide an introduction
to formal mathematics and thereby lower the barrier of entrance to the field.
So far, we have only conducted pilot experiments with
students of mathematics and computer science, an exemplary training session for
working with relations can be found
online.\footnote{\url{https://elfe-prover.org/tutorial}} It remains to be
investigated more thoroughly if and how \textsc{Elfe} can be used to teach
(formal) mathematics to young undergraduates and high-school students.

\bibliographystyle{eptcs}
\bibliography{references}

\end{document}

%% file: texts/tarski.tex
Notation between: a-b-c.\\
Notation equidistant: a-b $\equiv$ c-d. \\
Axiom CongrRefl: for all a,b. a-b $\equiv$ b-a. \\
Axiom CongrIdent: for all a,b,c. a-b $\equiv$ c-c implies a = b. \\
Axiom CongrTrans: for all a,b,p,q,r,s. a-b $\equiv$ p-q and a-b $\equiv$ r-s implies p-q $\equiv$ r-s.  \\
Axiom SegmentConstr: for all a,b,c,d. exists e. b-e $\equiv$ c-d and a-b-e. \\
Axiom FiveSegment: for all a,b,c,d,a',b',c',d'.  
   (a-b-c and a'-b'-c' and a-b $\equiv$ a'-b' and  b-c $\equiv$ b'-c' and a-d $\equiv$ a'-d' and b-d $\equiv$ b'-d' and not a = b) implies c-d $\equiv$ c'-d'. \\
Axiom BetwIdent: for all a,b. a-b-a implies a = b.\\
Axiom Pasch: for all a,b,c,p,q. a-p-c and b-q-c implies exists x. p-x-b and q-x-a. \\
Axiom LowerDim: exists a,b,c. not a-b-c and not b-c-a and not c-a-b. \\
Axiom Euclid: for all a,b,c,d,t. exists x,y. \\
\ind (a-d-t and b-d-c and not a = d) implies (a-b-x and a-c-y and x-t-y).

%% file: texts/geodefs.tex
Definition DefCol: for all a,b,c. col(a,b,c) iff a-b-c or b-c-a or c-a-b.\\
Definition DefMidpoint: for all a,b,m. \\ \ind midpoint(m,a,b) iff a-m-b and a-m $\equiv$ m-b.\\
Definition DefCoplanar: for all a,b,c,d. coplanar(a,b,c,d) iff exists x. \\
\ind (col(a,b,x) and col(c,d,x)) or  (col(a,c,x) and col(b,d,x)) or (col(a,d,x) and col(b,c,x)).\\
Notation parstr: a-b$|-|$c-d.\\
Definition DefParallelStrict: for all a,b,c,d. a-b$|-|$c-d iff \\ \ind (a $\neq$ b and c $\neq$ d and coplanar(a,b,c,d) and not exists x. col(x,a,b) and col(x,c,d)).\\
Notation parallel: a-b$||$c-d.\\
Definition DefParallel: for all a,b,c,d. a-b$||$c-d iff \\ \ind a-b$|-|$c-d or (a $\neq$ b and c $\neq$ d and col(a,c,d) and col(b,c,d)).